\documentstyle[epsf,graphics,epsfig,times]{mn}

\def\simgt{\mathrel{\lower0.6ex\hbox{$\buildrel {\textstyle >}
 \over {\scriptstyle \sim}$}}}
\def\simlt{\mathrel{\lower0.6ex\hbox{$\buildrel {\textstyle <}
 \over {\scriptstyle \sim}$}}}
\def\hompc{\,h\,{\rm Mpc}^{-1}}
\def\k{\mbox{\boldmath $k$}}
\def\r{\mbox{\boldmath $r$}}
\def\be{\begin{equation}}
\def\ee{\end{equation}}
\def\bm{\begin{displaymath}}
\def\em{\end{displaymath}}
\def\llangle{\left\langle}
\def\rrangle{\right\rangle}

\title[Fourier analysis of luminosity-dependent galaxy clustering] 
{Fourier analysis of luminosity-dependent galaxy clustering}

\author[Will J.\ Percival et al.]{
\parbox[t]{\textwidth}{
Will J.\ Percival$^1$,
Licia Verde$^{2,3}$,
John A.\ Peacock$^1$}
\vspace*{6pt} \\
$^1$Institute for Astronomy, University of Edinburgh, Royal Observatory, 
        Blackford Hill, Edinburgh EH9 3HJ, UK \\
$^2$Dept. Astrophysical sciences, Princeton University, Ivy Lane, 
	Princeton, NJ 08540, USA\\
$^3$ Department of Physics \& Astronomy, University of Pennsylvania, 
	209 South 33rd Street Philadelphia, PA 19104, USA \\
}
\date{Submitted for publication in MNRAS}
\begin{document}
\maketitle

\begin{abstract}
We extend the Fourier transform based method for the analysis of
galaxy redshift surveys of Feldman, Kaiser \& Peacock (1994: FKP) to
model luminosity-dependent clustering. In a magnitude limited survey,
galaxies at high redshift are more luminous on average than galaxies
at low redshift. Galaxy clustering is observed to increase with
luminosity, so the inferred density field is effectively
multiplied by an increasing function of radius. This has the potential
to distort the shape of the recovered power spectrum.  In this paper
we present an extension of the FKP analysis method to incorporate this
effect, and present revised optimal weights to maximize the precision
of such an analysis. The method is tested and its accuracy assessed
using mock catalogues of the 2-degree field galaxy redshift survey
(2dFGRS). We also show that the systematic effect caused by ignoring
luminosity-dependent bias was negligible for the initial analysis of
the 2dFGRS of Percival et al. (2001). However, future surveys,
sensitive to larger scales, or covering a wider range of galaxy
luminosities will benefit from this refined method.
\end{abstract}


\section{Introduction}  \label{sec:intro}

The shape of the matter power spectrum as traced by large-scale
structure is one of the cornerstones of modern cosmology. The shape is
dependent on the primordial fluctuations, on the total matter
density $\Omega_m$ through $\Omega_mh$ which controls the
matter-radiation horizon scale, and through the fraction of matter in
baryons $\Omega_b/\Omega_m$. Although these parameter combinations are
nearly degenerate, preliminary results from the 2-degree field galaxy
redshift survey (2dFGRS) contained sufficient signal to weakly break
this degeneracy (Percival et al. 2001: P01). The complete 2dFGRS, the
Sloan Digital Sky Survey (SDSS; York et al. 2000) and forthcoming
high-redshift surveys will refine these measures of the cosmological
matter content from large-scale structure observations.

High-quality measurements of the cosmic microwave background (CMB)
power spectrum (e.g. Bennett et al. 2003; Hinshaw et al. 2003) have
brought about a new era of high-precision cosmology, in which the
analysis of large-scale structure (LSS) data assumes even greater
importance.  Many of the degeneracies between cosmological parameters
intrinsic to CMB data are broken by the addition of constraints from
LSS data (e.g. Efstathiou et al. 2002; Percival et al. 2002; Spergel
et al. 2003; Verde et al. 2003). Thus, the scientific potential of
both data sets can be greatly enhanced through combination. However,
the physics and instrumental effects involved in the interpretation of
LSS data are more complicated and less well understood than for the
CMB data. To use LSS reliably in the context of high-precision
cosmology, it is therefore important to consider the survey analysis
method and any possible systematic problems in detail.

There are a number of difficulties and potential systematic effects
associated with recovering the shape of the matter power spectrum from
that of the galaxies. Distances to galaxies are derived from
redshifts, and peculiar velocities induce systematic distortions in
the recovered power. Although this effect is important and potentially
carries cosmological information (e.g, Peacock et al. 2001; Hawkins et
al. 2003; see Hamilton 1998 for a review), we will not address it
here, but we will consider a more subtle effect. Galaxies are biased
tracers of the matter distribution: the relation between the galaxy
and mass density fields is probably both nonlinear and stochastic to
some extent (e.g. Dekel \& Lahav 1999), so that the power spectra of
galaxies and mass differ in general. If we define scale-dependent bias
via
\be
 P_g(k)=b^2(k) P_m(k),
    \label{eq:biasdef}
\ee
where subscripts $m$ and $g$ denote matter and galaxies respectively,
there is no guarantee that $b(k)$ will be a constant. The best we can
hope for is that there will be a `linear response' limit, in which
$b(k)$ tends to a constant $b_{\rm lin}$ on large scales.  Although it
is easy to invent artificial models in which this is not true, the
concept of linear bias does hold for many bias models -- and in
particular for the most detailed attempts to include all the physics
of galaxy formation (e.g. Benson et al. 2000).  Such simulations can
also give a realistic idea of the scales on which the linear-bias
assumption breaks down, and we assume for the purposes of this paper
that linear bias is true (or that small deviations from
scale-independence can be corrected for). Nevertheless, even in the
linear-bias limit, bias can complicate the analysis of LSS data,
because the value of $b_{\rm lin}$ will be different for different
classes of galaxy. The purpose of this paper is to investigate the
extent to which this can affect the recovered power spectrum shape.

The fact that galaxies selected in different ways have different
clustering properties has been known for some time (e.g. Davis \&
Geller 1976; Peacock \& Dodds 1994; Seaborne et al. 1999). For the
2dFGRS galaxies, although the average bias is close to unity (Lahav et
al. 2002; Verde et al. 2002), the bias is dependent on galaxy
luminosity (Norberg et al. 2001; 2002; Zehavi et al. 2002 find a very
similar dependence for SDSS galaxies), with
\be
  \llangle \frac{b(L)}{b(L_*)}\rrangle
    = 0.85 + 0.15 \frac{L}{L_*},
    \label{eq:bobstar}
\ee
where the bias $b(L)$ is assumed to be a simple function of galaxy
luminosity and $L_*$ is defined such that $M_{b_{\rm J}}-5\log_{10}h=-19.7$
(Norberg et al. 2001).

In a magnitude limited survey, galaxies of different luminosity are
probed at different radii: at high redshift, galaxies more luminous
than average (and with lower number density) dominate the sample,
while at low redshift galaxies less luminous than average (and with
higher number density) dominate. As a consequence, the power spectrum
at large scales is measured preferentially from galaxies more luminous
than average, and on small scales from galaxies less luminous than
average. Without correction, luminosity-dependent bias would therefore
distort the shape of the recovered galaxy power spectrum from that of
the matter power spectrum (see Tegmark et al. 2003).

In this paper we consider how best to estimate the shape of the
underlying matter power spectrum given a sample of galaxies that have
a clustering amplitude dependent on galaxy properties. Specifically,
in Section~\ref{sec:lumbias} we consider a set of galaxies with a
simple luminosity-dependent bias such that, on the scales of interest
$P_g(k)=b^2(L) P_m(k)$. For generality, the expected bias need not be
simply a function of luminosity; all that is required is that the bias
can be predicted from some combination of the properties of each
galaxy. The results of this paper are therefore relevant to more
general problems where we have a mixed set of galaxies tracing the
same density field. For instance the method would be applicable to
type-dependent clustering or a survey covering a large redshift range
where the linear evolution of the matter power spectrum and evolution
in the bias of the objects selected were important.

Previous studies have already extended the calculation of optimal
weights presented by FKP. Hamilton (1997) considered the effect of the
window function which the FKP method assumes to be negligible. More
recently, Yamamoto (2003) attempted to derive optimal weights
including the effect of both redshift-space distortions and the
light-cone effect, which, as we will illustrate later, has a similar
net effect as luminosity dependent bias. In this paper, the work of
FKP is extended in a different context to the work of Yamamoto (2003),
and we consider the systematic effect of luminosity-dependent bias on
the recovered power in addition to providing optimal
weights. Interestingly, our derived optimal weights, following an
independent calculation, differ from those of Yamamoto (2003) and a
comparison is given in Section~\ref{sec:discussion}.

The layout of this paper is as follows. First, in
Section~\ref{sec:fkp_intro} we present a generalization of the power
spectrum analysis method introduced by Feldman, Kaiser \& Peacock
(1994: FKP), extended to correct for galaxy luminosity dependent
bias. We then follow the FKP derivation leading to revised optimal
weights in Section~\ref{sec:optimise}, in the limit of a large survey
volume. The method and weights are tested in Section~\ref{sec:mocks}
using simple mock catalogues with a window function similar to that of
the 2dFGRS, and the relevance for the analysis of P01 is considered in
Section~\ref{sec:2dFGRS}. We conclude in Section~\ref{sec:discussion}.

\section{Analysis of a mixed luminosity sample} \label{sec:lumbias}

In this Section we extend the method of FKP to cover a sample of
galaxies with different large-scale biases. In order to make
the description transparent, we adopt the notation of FKP, and
refer extensively to this paper. In particular, we adopt the
Fourier transform convention of FKP, equivalent to that of Peebles
(1980) with $V=1$.

\subsection{Method}  \label{sec:fkp_intro}
Let us consider a set of galaxies of luminosity $L$ forming a Poisson
sampling of a linearly biased density field 
\be
1 + \delta_g(\r) = 1+b(\r,L)\delta(\r).
\ee
Here $b(\r,L)$ is the linear bias of galaxies of luminosity
$L$ at position $\r$. We have generalized the idea of bias
being a function of $L$ only to allow for e.g. slow evolution
of $b(L)$ within the survey volume. 
The probability that volume element $\delta V$ contains a
galaxy of luminosity $L$ is given by
\be
  {\rm Prob}(\delta V,L) = \delta V \bar{n}(\r,L)[1+b(\r,L)\delta(\r)],
\ee
where $\bar{n}(\r,L)$ is the mean expected number density of galaxies
at $\r$ with luminosity $L$. The power spectrum and correlation
function form a Fourier pair with $\xi(\r)=\xi(r)=\llangle
\delta(\r')\delta(\r'+\r)\rrangle$, and
\be
  P(\k) = P(k) = \int d^3r\; \xi(\r)e^{i\k\cdot\r}.
\ee

Here and hereafter $P(k)$ denotes the power spectrum of the underlying
matter fluctuation field $\delta$. If, rather than considering the bias
of each galaxy, we had instead considered the ratio of the bias to
that of a particular galaxy type, then we simply have to redefine
$P(k)$ as the power spectrum of that galaxy type. Replacing $b(\r,L)$
by $b(\r,L)/b(\r,L*)$ would mean that $P(k)$ would have to be
redefined as the power spectrum of $L_*$ galaxies so, for the 2dFGRS
galaxies, we could use Eq.~\ref{eq:bobstar} to recover the power
spectrum of $L_*$ galaxies.

Following the FKP approach, we also consider a synthetic catalogue
with the same radial and angular sampling (selection function), and
luminosity distribution as the galaxy catalogue but with no
correlations. This synthetic catalogue will be used (Eq.~\ref{eq:Fr})
to convert from a galaxy density field to a galaxy overdensity
field. We will multiply objects in both the galaxy and synthetic
catalogue by $w(\r,L)/b(\r,L)$ -- here the split distinguishes the
bias correction and additional weighting needed to optimize the
precision of the power estimate. For clarity in the following text, we
only use the ``weight'' to refer to $w(\r,L)$, not $w(\r,L)/b(\r,L)$.

The weighted galaxy fluctuation field is defined as
\be
  F(\r)\equiv\frac{1}{N}\int dL\,\frac{w(\r,L)}{b(\r,L)}
    \left[n_g(\r,L)-\alpha n_s(\r,L)\right]
    \label{eq:Fr}
\ee
where $n_g(\r,L)=\sum_j\delta(\r-\r_j)\delta(L-L_j)$, with $\r_j$ being the
location of the $j$th galaxy of luminosity $L_j$, and $n_s(\r,L)$ is
defined similarly for the synthetic catalogue. Here $\alpha$ is a
constant that matches the two catalogues (see
Section~\ref{sec:alpha}), and $N$ is a normalization constant defined by
\be
  N=\left\{\int d^3r\left[\int dL\,
    \bar{n}(\r,L)w(\r,L)\right]^2\right\}^{1/2}.
  \label{eq:N}
\ee
Following appendix~A of FKP, the two-point functions of $n_g,n_s$ are
\bm
  \llangle n_g(\r,L) n_g(\r',L') \rrangle = 
\em
\bm
  \hspace{1cm}
  \bar{n}(\r,L)\bar{n}(\r',L')\left[1+b(\r,L)b(\r',L')\xi(\r-\r')\right]
\em
\be
  \hspace{1cm}
    + \bar{n}(\r,L)\delta(\r-\r')\delta(L-L'),
  \label{eq:exp1}
\ee
\bm
  \llangle n_s(\r,L) n_s(\r',L') \rrangle = 
    \alpha^{-2}\bar{n}(\r,L)\bar{n}(\r',L')
\em
\be
  \hspace{1cm}
    + \alpha^{-1}\bar{n}(\r,L)\delta(\r-\r')\delta(L-L'),
  \label{eq:exp2}
\ee
\be
  \llangle n_g(\r,L) n_s(\r',L') \rrangle = 
    \alpha^{-1}\bar{n}(\r,L)\bar{n}(\r',L'),
  \label{eq:exp3}
\ee
where we have explicitly included the bias caused by the different
galaxy luminosities. The expected value $\llangle F(\r)F(\r')\rrangle$ is
therefore given by
\be
  \llangle F(\r)F(\r')\rrangle = G(\r)G(\r')\xi(\r-\r')
    + \xi_{\rm shot}(\r,\r'),
    \label{eq:expFrFr}
\ee
where
\be
  G(\r) = \frac{1}{N}\int dL\,\bar{n}(\r,L)w(\r,L),
\ee
and
\be
  \xi_{\rm shot}(\r,\r') = 
    \frac{1+\alpha}{N^2}\int dL\,\bar{n}(\r,L)\frac{w^2(\r,L)}{b^2(\r,L)}
    \delta(\r-\r').
  \label{eq:xishot}
\ee
The first term in Eq.~\ref{eq:expFrFr} corresponds to the multiplication
of the correlation function by the window. The second is the shot
noise term. Switching to Fourier space we have, as in FKP, that 
\be
  \llangle |F(\k)|^2 \rrangle
    = \int\frac{d^3k'}{(2\pi)^3}P(\k')|G(\k-\k')|^2 + P_{\rm shot}.
 \label{eq:expFkFk}
\ee
The multiplication in real space (the first term in
Eq.~\ref{eq:expFrFr}) has become a convolution in Fourier space (the
first term in Eq.~\ref{eq:expFkFk}),
but now
\be
  G(\k) = \frac{1}{N}\int d^3r \int dL\,
    \bar{n}(\r,L)w(\r,L)e^{i\k\cdot\r},
  \label{eq:Gk}
\ee
and
\be
  P_{\rm shot} = 
    \frac{1+\alpha}{N^2}\int d^3r  \int dL\,
    \bar{n}(\r,L)\frac{w^2(\r,L)}{b^2(\r,L)}.
  \label{eq:Pshot}
\ee
Ignoring the bias changes between galaxies with different luminosities
reduces $G(\k)$ and $P_{\rm shot}$ to the FKP expressions.

The factorization of the multiplicative factor in Eq.~\ref{eq:Fr} is
now clear: although we multiply by $w(\r,L)/b(\r,L)$, the window
function shape (Eq.~\ref{eq:Gk}) and normalization (Eq.~\ref{eq:N})
are only dependent on the weights $w(\r,L)$: the additional
$1/b(\r,L)$ factor simply corrects the measured galaxy fluctuations to
an estimate of the matter fluctuations, and does not affect the
window.

Our estimator of $P(\k)$ convolved with the window function is given
by $\hat{P}(\k)=|F(\k)|^2-P_{\rm shot}$. Averaging over a shell in
$k$-space, gives our final estimator of the convolved power $P(k)$,
$\hat{P}(k)$,
 \be
  \hat{P}(k)\equiv\frac{1}{V_k}\int_{V_k}d^3k'\hat{P}(\k'),
  \label{eq:estPk}
\ee
where $V_k$ is the volume of the shell. Ignoring redshift-space
distortions, $\hat{P}(k)$ is the true power spectrum convolved with a
spherically-averaged window, which can be computed by spherically
averaging $|G(\k)|^2$, with $G(\k)$ given by Eq.~\ref{eq:Gk}.
Redshift-space distortions mean that the interpretation of
Eq.~\ref{eq:estPk} is actually more complicated, and $\hat{P}(k)$
depends on the full survey window. In the spirit of generalizing FKP,
such effects are not included here; a separate study of this issue
would nevertheless be of interest.

We can now contrast the FKP procedure with the exact analysis of
luminosity-dependent clustering, as expressed in
Eqns.~\ref{eq:expFkFk} --~\ref{eq:Pshot}. FKP used
luminosity-independent weights, equivalent to $w(\r,L) = b(\r,L) \,
w_{\rm\scriptscriptstyle FKP}(\r)$ in our current notation. The
resulting density fluctuation field (Eq.~\ref{eq:Fr}) differs from the
corresponding quantity in FKP, because the normalization factor, $N$,
contains $w({\bf r},L)$.  We therefore have $F_{\rm\scriptscriptstyle
FKP}(\r) = b_{\rm eff} F(\r)$, where

\be
  b_{\rm eff}^2 = \frac
    {\int d^3r \left[\int dL\,\bar n(\r,L)\, b(\r,L)\, 
    w_{\rm\scriptscriptstyle FKP}(\r) \right]^2}
    {\int d^3r \left[\int dL\,\bar n(\r,L)\,
    w_{\rm\scriptscriptstyle FKP}(\r) \right]^2}
\ee
This effective bias is potentially more serious than just a shift in
the overall normalization, since $w_{\rm\scriptscriptstyle FKP} =
1/[1+\bar n P(k)]$ is a function of wavenumber: a different $F(\r)$ is
to be transformed for each $k$ value. This means that $b_{\rm eff}$ is
also a function of $k$, and the shape of the recovered spectrum is
systematically altered.  However, most analyses (including P01) have
not taken the FKP mantra to this extreme, and have in practice assumed
a single value for $P(k)$ at all wavenumbers.  Therefore, the P01
analysis is not affected by this potential difficulty.

Even so, there is a second difference between our results and FKP,
which may be seen in Eq. 15. The derived power spectrum is an estimate
of the true power convolved with a window function $|G(k)|^2$, and the
correct form for this window differs from the corresponding function
in the FKP analysis.  As shown in P01, this convolution significantly
changes the shape of the recovered power spectrum, and so we are led
to ask whether the change in the window is important; we show below in
Section 3.4 that it is not.

Finally, although we are thus able to identify potential systematic
errors that can arise from application of the FKP procedure to
luminosity-dependent clustering, this does not address the issue of
optimality. In fact, the original FKP weight is not optimal in general
-- nor is the simple bias-corrected multiple of it. A derivation of
the correct optimal weight (in the usual limit of a narrow $k$-space
window) is presented in Section~\ref{sec:optimise}.

\subsection{Statistical fluctuations in the convolved 3D power}  
  \label{sec:fluc}

We now proceed to analyse the error in our power estimator, as in FKP
section 2.2. We assume that the Fourier components $F(\k)$ are
Gaussian-distributed, so the variance in our estimate of the convolved
power $\llangle \delta\hat{P}(\k)\delta\hat{P}(\k')\rrangle = \left|\llangle
F(\k)F^*(\k')\rrangle\right|^2$ (FKP Appendix~B). On the large-scales of
interest this is a good approximation in the limit of a compact window
function. Following the same steps as in FKP section~2.2, our estimate
of the mean square fluctuations in the recovered convolved power
$\llangle F(\k)F^*(\k+\delta\k)\rrangle$ is given by
\be
  \llangle F(\k)F^*(\k+\delta\k)\rrangle \simeq 
    P(\k)Q(\delta\k)+S(\delta\k),
\ee
where
\be
  Q(\k) = \frac{1}{N^2}\int d^3r 
    \left[\int dL\,\bar{n}(\r,L)w(\r,L)\right]^2
    e^{i\k\cdot\r},
  \label{eq:Qk}
\ee
and
\be
  S(\k) = \frac{1+\alpha}{N^2}\int d^3r 
    \int dL\,\bar{n}(\r,L)\frac{w^2(\r,L)}{b^2(\r,L)}.
    e^{i\k\cdot\r}.
  \label{eq:Sk}
\ee
This leads to an estimate of the error in the convolved power 
\be
  \llangle \delta\hat{P}(\k)\delta\hat{P}(\k')\rrangle =
    |P(\k)Q(\delta\k)+S(\delta\k)|^2.
\ee

\subsection{Optimal Weighting}  \label{sec:optimise}

Optimal weights can now be derived for such an
analysis following a direct extrapolation of FKP section
2.3.

We assume that the window function is compact in $k$-space compared
with the scales of interest; this will be discussed in more detail in
Section~\ref{sec:discussion}.  The mean square fluctuation in our
estimate of the power is
\begin{eqnarray}
  \sigma_P^2(k) & \equiv & 
    \llangle\left[\hat{P}(k)-P(k)\right]^2\rrangle\\
  & = & \frac{1}{V_k^2}\int_{V_k}d^3k\int_{V_k}d^3k'
    \llangle\delta\hat{P}(\k)\delta\hat{P}(\k')\rrangle.
\end{eqnarray}

If the shell over which we average has a width that is large compared
to the effective width of $Q(\k)$, but small compared with
the variation in $P(k)$, then this double integral reduces to
\be
  \sigma_P^2(k) \simeq
    \frac{1}{V_k}\int d^3k'\left|P(k)Q(\k')+S(\k')\right|^2.
\ee
Using the definitions of $Q(\k)$ and $S(\k)$ from Eqns.~\ref{eq:Qk}
\&~\ref{eq:Sk}, and Parseval's theorem, the fractional variance in the
power is
\bm
  \frac{\sigma_P^2(k)}{P^2(k)} = \frac{(2\pi)^3}{V_kN^4}
    \int d^3r\left\{\left[\int dL\,\bar{n}(\r,L)w(\r,L)\right]^2
    \right.
\em
\be
  \hspace{1cm} \left. + 
    \left[\int dL\,\bar{n}(\r,L)\frac{w^2(\r,L)}{b^2(\r,L)}\right]
    \frac{1}{P(k)}\right\}^2.
  \label{eq:fracvar}
\ee

We wish to find $w(\r,L)$ that minimize Eq.~\ref{eq:fracvar}. Without
loss of generality, we consider a small variation in the weights
$w(\r,L)\to w(\r,L)+\delta(L-L')\delta w'(\r)$. To keep the equations
concise, we set $w\equiv w(\r,L)$, $\bar{n}\equiv\bar{n}(\r,L)$,
$b\equiv b(\r,L)$ and $P\equiv P(k)$. Similarly, $w'\equiv w(\r,L')$,
$\bar{n}'\equiv\bar{n}(\r,L')$ and $b'\equiv b(\r,L')$. Requiring that
$\sigma_P^2(k)$ be stationary with respect to arbitrary variations
$\delta w'(\r)$ gives that
\bm
  \frac{\int d^3r 
    \left[ \left(\int dL\,\bar{n}w\right)^2
    + \left(\int dL\,\frac{\bar{n}w^2}{Pb^2}\right)\right]
    \left[\bar{n}'\left(\int dL\,\bar{n}w\right)
    + \frac{\bar{n}'w'}{Pb'^2}\right]\delta w'}
    {\int d^3r 
    \left[ \left(\int dL\,\bar{n}w\right)^2
    + \left(\int dL\,\frac{\bar{n}w^2}{Pb^2}\right)\right]^2}
\em
\be
  \hspace{1cm}
    = \frac{\int d^3r \, \bar{n}'
    \left(\int dL\,\bar{n}w\right)\delta w'}
    {\int d^3r \left(\int dL\,\bar{n}w\right)^2}.
    \label{eq:big}
\ee
The non-trivial solution to this Equation is
\be
  w(\r,L') = 
    \frac{b^2(\r,L')P(k)}{1+\int dL\,\bar{n}(\r,L)b^2(\r,L)P(k)},
    \label{eq:w}
\ee
which is the principal result of this paper.

This formula was found by considering initially the case of two sets
of galaxies with different luminosities, so the integral is replaced
by a sum over the two subsets. The resulting equations were then
solved with the aid of {\sc mathematica}. The form of the 2-class
solution suggested a conjecture for the general solution with a
countable number of sets of galaxies, and thus for a set of galaxies
with a continuous distribution of luminosities. The conjecture was
readily verified by direct substitution.  It is straightforward to see
that this reduces to the formula of FKP (their equation 2.3.4) for a
sample of galaxies at a single luminosity.

Eq.~\ref{eq:w} shows that $w(\r,L)$ depends not only on the expected
bias of galaxies of that luminosity $b(\r,L)$, but additionally on the
bias of galaxies of all luminosities at this location through $\int
dL\,\bar{n}(\r,L)b^2(\r,L)$. This follows because the balance between
shot noise and cosmic variance, which is at the heart of the
derivation of the optimal weights, depends on what we learn from all
galaxies whatever their luminosity. It is also interesting to note
that, while the contribution of galaxies to the overdensity estimate
appears to be inversely proportional to their bias (the $1/b(\r,L)$
factor in Eq.~\ref{eq:Fr}), optimizing the weight actually means that
the net contribution of these galaxies is increased -- they contain
the most signal-to-noise because of their strong clustering.

\subsection{Choice of $\alpha$}  \label{sec:alpha}

The value of $\alpha$ in Eq.~\ref{eq:Fr} sets the expected number of
galaxies for a particular survey. In this Section we discuss how this
can be set given no information other than the survey itself.

Suppose that we had used an incorrect value of $\alpha$ in
Eq.~\ref{eq:Fr}, so that this Equation became
\be
  F(\r)\equiv\frac{1}{N}\int dL\,\frac{w(\r,L)}{b(\r,L)}
    \left[n_g(\r,L)-(\alpha-\beta)n_s(\r,L)\right],
\ee
then the effect of the additional component is to introduce an
additional term 
\be
  \frac{\beta^2}{\alpha^2}\int\frac{d^3k}{(2\pi)^3}|G(\k)|^2
\ee
to the right hand side of Eq.~\ref{eq:expFkFk}. Turning this argument
around, we see that not knowing the true value of $\alpha$ is
equivalent to adding a multiple of the Fourier transformed window
function to our power estimate. In general, the average weighted
galaxy density has to be derived from the survey itself, and it is
therefore impossible to know the true offset between the number of
galaxies observed and that expected. In this situation, the only
sensible thing to do is to set
\be
  \alpha = \frac{\int d^3r \int dL\,\frac{w(\r,L)}{b(\r,L)}n_g(\r,L)}
    {\int d^3r \int dL\,\frac{w(\r,L)}{b(\r,L)}n_s(\r,L)},
\ee
so that the average weighted overdensity is artificially set to
zero. This self-normalization forces $P(0)=0$ and results in a deficit
in the estimated power equivalent to subtracting a scaled copy of the
window function, centered on $k=0$ (Peacock \& Nicholson 1991). In the
limit of no window, the effect of this self-normalization would be to
remove a delta function located at $k=0$ such that $P(0)=0$.  In
effect, this procedure evades the possibility of noise in our estimate
of large-scale power owing to the uncertainty in $\bar n$, at the
expense of systematic damping of the large-scale signal (cf. the
approach of Tadros \& Efstathiou 1995).

\subsection{A note on the convention adopted}

Instead of allowing for a continuous distribution of galaxy
luminosities and corresponding biases, an alternative approach would have
been to consider a countable number of sets of galaxies with different
luminosities. In this case, Eq.~\ref{eq:Fr} would have become
\be
  F(\r)\equiv\frac{1}{N}\sum_i\frac{w(\r,L_i)}{b(\r,L_i)}
    \left[n_g(\r,L_i)-\alpha n_s(\r,L_i)\right]
    \label{eq:Fr2}
\ee
where $n_g(\r,L_i)=\sum_j\delta(\r-\r_j)$ now gives the density of
galaxies within subset $i$, the subset containing those galaxies of
luminosity $L_i$. The analysis and derivation of optimal weights
described in this paper follows exactly as for the continuous case,
except that we obviously need to replace the integral over luminosity
with a sum over the subsets of galaxies throughout.

\section{Application to mock catalogues}  \label{sec:mocks}

Having introduced the necessary formalism, we now illustrate the
revised FKP method using simple artificial surveys. These artificial
surveys were designed to approximate the selection function and
luminosity-dependent bias of the 2dFGRS galaxies analysed by P01.

\subsection{Selection function}  \label{sec:selection}

The selection function is approximately matched to the part-complete
2dFGRS survey (Colless et al. 2001) as analysed in P01. The redshift
distribution used for the galaxies is given by
\be 
  f(z) = z^{1.25}\exp\left[-\left(\frac{z}{0.13}\right)^{2.2}\right],
\ee 
and limited to $0<z<0.25$. 

This radial selection function corresponds to the angle-averaged
selection function of the survey as analyzed in P01 including both the
NGP, SGP and random fields. However the selection function does not
exactly match that of P01 as it does not include the varying survey
depth of the 2dFGRS sample (see Colless et al. 2001 for details). As
we simply wish to compare and contrast our revised analysis method and
the effect of not taking into consideration luminosity-dependent bias,
this difference is not important. The expected number of galaxies in
each catalogue was set at $140\,000$, although the actual number varies
through Poisson sampling (see Section~\ref{sec:mockcreate}).

In the following analysis we use a $512\times512\times256$ grid that
just covers the survey region: given the survey geometry, the
rectangular nature of this grid yields roughly similar sampling, and
therefore Nyquist frequencies, in each Cartesian direction.

\subsection{Creating the mock catalogues}  \label{sec:mockcreate}

In this Section we describe how the mock galaxy catalogues, drawn from
lognormal random fields, were generated (this closely follows the
method described in Coles \& Jones 1991). Lognormal random fields were
used for convenience because they obey the physical limit
$\delta(\r)>-1\;\forall\r$, and approximate the present-day non-linear
fluctuation field.

The underlying matter power spectrum was chosen to be a $\Lambda$CDM
model calculated from the transfer function fitting formulae of
Eisenstein \& Hu (1999) coupled with the non-linear fitting formulae
of Smith et al. (2002). The following cosmological parameter values
were set: $\Omega_{\rm m}=0.3$, $\Omega_{\rm b}/\Omega_{\rm m}=0.15$,
$\Omega_\Lambda=0.7$, $h=0.7$, $n_{\rm s}=1$ (consistent with the
recent WMAP results -- Spergel et al. 2003). The normalization of the
power spectrum (chosen to be $\sigma_8=1$) is unimportant for our
analysis, which focuses on the relative change in the recovered power
with various weighting schemes rather than the absolute value. In
addition, we will use Eq.~\ref{eq:bobstar} to determine bias relative
to $L_*$ galaxies rather than the absolute bias, so the power spectrum
normalization chosen should correspond to that of $L_*$ galaxies.

This power spectrum was determined on the grid described in
Section~\ref{sec:selection}. In order to avoid a sharp cut-off in the
power, we introduce a smooth turn over at 0.1 times the minimum
Nyquist frequency, $\nu_{\rm Ny}$, which cuts the input power at
$0.25\nu_{\rm Ny}$. This does not affect any of our conclusions as we
are only interested in the large-scale power $k<0.15\hompc$. The input
power was inverse Fourier transformed to obtain the correlation
function of the lognormal field required. The covariance of the
Gaussian field, required to generate the lognormal field, is then
obtained from $\xi_{\rm G}(r)=\ln[1+\xi_{\rm LN}(r)]$, and this was
converted back to the power. A Gaussian density field $\delta_{\rm G}$
was then generated on the grid with this power spectrum, and the
corresponding lognormal field was calculated, given by $\delta_{\rm
LN}=\exp(\delta_{\rm G}-\sigma_{\rm G}^2/2)-1$, where $\sigma_{\rm
G}^2$ is the variance of the Gaussian density field.

The lognormal density field was then used to create a catalogue that
matches the luminosity-dependent clustering of the 2dFGRS sample. To
facilitate this process we made a number of simplifications. First, we
used the same grid both to create the catalogue and to estimate the
power spectrum. There was therefore no need to create a full
catalogue; instead, we can determine a catalogue already sampled on
the grid. The additive nature of the Poisson distribution meant that
the total number of galaxies at each grid point could be calculated by
drawing a random Poisson variable with mean given by the selection
function multiplied by the lognormal field and the mean expected bias
at that grid point. These galaxies then need to be assigned
luminosities. As we simply wish to use the mock catalogues to test the
analysis method, we choose to only model the net effect, which is that
the effective bias at a survey location depends on the distance from
the observer. We therefore assume that the galaxies at each grid point
all have the same luminosity, given by the average luminosity of the
2dFGRS sample at that distance, and that they have the same average
expected bias. This approach can be interpreted as a special case
where we are modeling power spectrum evolution along the line-of-sight
rather than luminosity-dependent clustering.

We now describe how we calculated the biased lognormal density
field. For the 2dFGRS sample of P01, the mean expected bias as a
function of redshift is well fit by the formula
\be 
  \llangle b(\r)\rrangle = \int dL\,\bar{n}(\r,L)b(\r,L)
    = 0.85 + 6z^{1.75},
  \label{eq:bias_radial}
\ee
where $z$ is the redshift corresponding to a given radius.

There are a number of ways that one could apply this formula to create
a biased density field. Perhaps the most rigorous method would be to
create a number of lognormal fields with the same phases and different
amplitudes, each corrected from the desired clustering strength as
detailed above. We should then use the field corresponding to the
desired power amplitude at each grid point. However, this would have
been computationally very expensive. In the linear regime of interest,
the fields are all of small amplitude and the correction between
Gaussian and lognormal fields is very small. We therefore created a
single lognormal field corresponding to the largest mean expected bias
in the 2dFGRS sample $b(L)\sim 1.4b(L_*)$, and multiplied this field
by the expected mean bias given by Eq.~\ref{eq:bias_radial} divided by
$1.4$. By choosing to create the field for the largest mean bias we
ensure that $\delta_{\rm LN}(\r)>-1\;\forall\r$. Alternatives would have
been to bias either the Gaussian field $\delta_{\rm G}\to
b(L)\delta_{\rm G}$, or to set $(1+\delta_{\rm LN})\to(1+\delta_{\rm
LN})^{b(L)}$. This choice is not significant, and all options result
in fields with $\delta_{\rm LN}(\r)>-1\;\forall\r$, and approximately the
correct varying clustering strength.

\subsection{Analysis of mock catalogues}  \label{sec:mockmethod}

In order to test our derivation of optimal weights, we consider three
weighting schemes. First we consider the weights derived to be optimal
for a narrow $k$-space window 
\be
  w_1 = \frac{b^2(\r,L)P(k)}{1+\int dL\,\bar{n}(\r,L)b^2(\r,L)P(k)}.
    \label{eq:w1}
\ee

The second weighting scheme that we consider is designed to highlight how
luminosity-dependent bias affects the results of P01, and is given by
\be
  w_2 = \frac{b(\r,L)P(k)}{1+\int dL\,\bar{n}(\r,L)b^2(\r,L)P(k)}.
    \label{eq:w2}
\ee
The $b(\r,L)$ term in the numerator cancels the $1/b(\r,L)$ factor in
Eq.~\ref{eq:Fr} applied to each galaxy to correct the differing
clustering strengths. Using these weights is therefore equivalent to
the original FKP scheme, except that there is an additional
$b^2(\r,L)$ in the denominator. This form was chosen rather than the
exact FKP weight because this additional term is simply equivalent to
varying $P(k)$ as a function of $\r$, and keeping the denominator
fixed for all three weights enables the effect of the bias term in the
numerator to be more easily understood.

In addition we consider weights that apply no distinction between
galaxies with different expected biases
\be
  w_3 = \frac{P(k)}{1+\int dL\,\bar{n}(\r,L)b^2(\r,L)P(k)}.
    \label{eq:w3}
\ee

For the analysis presented in the next Section, we assumed a fixed
value of $P(k)=5000\,h^3\,{\rm Mpc}^{-3}$ for all of the weighting
schemes. The squared window functions $|G(\k)|^2$ were calculated as
in Eq.~\ref{eq:Gk}, and were spherically averaged as for the power
estimate (Eq.~\ref{eq:estPk}). The data were then fitted with a smooth
curve calculated using a Spline3 algorithm (Press et~al. 1992). This
curve was found to be a better fit to the shape of the recovered
window at $k>0.1\hompc$ than the fitting formula of P01. For
$k>0.6\hompc$, we extrapolate the window function as $k^{-4}$. This
extrapolation does not significantly affect the shape of the recovered
convolved power, although it does have a slight effect on the
normalization.

As we have shown in Eqns.~\ref{eq:Fr} \&~\ref{eq:Gk}, although the
galaxies were multiplied by the weight divided by the expected bias,
the window function is only dependent on the weight. In an FKP
analysis, such as P01, the weight assumed for the window function is
the same as the multiplicative factor in Eq.~\ref{eq:Fr},
$w(\r,L)/b(\r,L)$. For example, if weights $w_2$ were applied to the
galaxies in an FKP analysis of a catalogue with luminosity dependent
clustering, weights $w_3$ would be assumed for the window
function. Similarly, if weights $w_1$ were applied to the galaxies,
weights $w_2$ would be assumed for the window function. Because we fix
$P(k)$ in these weights as in P01, then the change in the shape of the
recovered power spectrum is only dependent on this change in the
window (see Section~\ref{sec:fkp_intro}). The importance of the
changing window on the shape of the power spectrum is considered in
the next Section where we present the recovered power spectra from our
mock catalogues calculated using weights $w_1$, $w_2$, and $w_3$.

Because we create the mock catalogues and power estimates on the same
grid, we know the selection function at each grid point exactly. We
can therefore easily subtract the mean galaxy density without having
to create a synthetic catalogue. In order to determine the shot-noise
level and the power normalization we require two integrals, given by
Eqns.~\ref{eq:Pshot}~\&~\ref{eq:N} respectively. As we are assuming
that the distribution of galaxy luminosities is a delta function at
each grid point, and we know the selection function at these points,
then it is relatively straightforward to calculate these two
quantities by numerically integrating over the grid used. In the
following we apply the method described in Section~\ref{sec:fkp_intro}
to estimate the power at $100$ points evenly spaced in
$0<k<0.128\pi\hompc$. This sampling is matched to that of P01.

\subsection{Results}  \label{sec:results}

\begin{figure}
  \setlength{\epsfxsize}{\columnwidth} 
    \centerline{\epsfbox{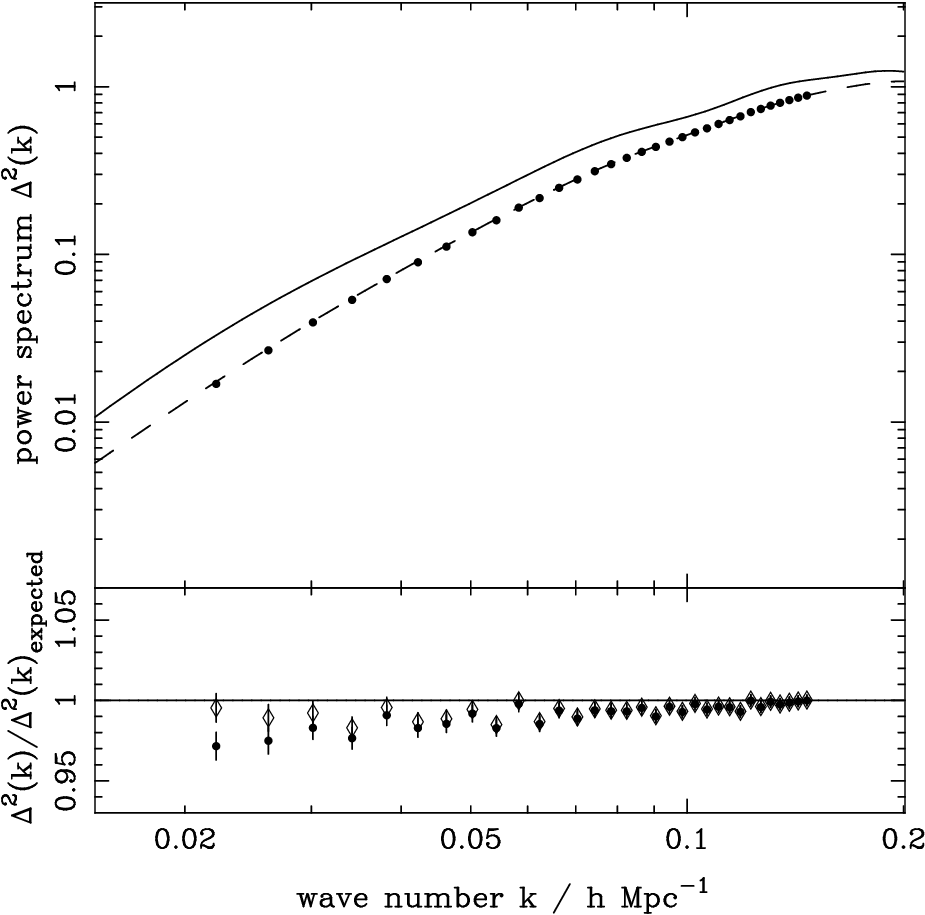}}

  \caption{Top panel: the average recovered power spectrum from $1000$
  mock catalogues calculated within a selection function designed to
  mimic the 2dFGRS sample analysed by P01 (solid circles). These data
  were weighted using weights $w_1$ given by Eq.~\ref{eq:w1}. For
  comparison we also plot the input power spectrum (solid line) and
  the power spectrum convolved with the fit to the spherically
  averaged Fourier window function (dashed line). In the lower panel,
  these data are compared with the models in more detail, and we plot
  the ratio of the average recovered power to the convolved model
  (solid circles with 1$\sigma$ errors). The open diamonds show the
  ratio of the data to the expected convolved power including
  correction for the self-normalization induced by matching the
  average galaxy density (see Section~\ref{sec:alpha} for
  details). \label{fig:Dsq_w1}}
\end{figure}

We have created $1000$ lognormal catalogues as described in
Section~\ref{sec:mockmethod}, and have analysed these catalogues using
the three different weighting schemes given in
Section~\ref{sec:mockmethod}. In this paper we only consider the
recovered power for $0.02<k<0.15\hompc$. This matches the region
considered in P01 and avoids the non-linear regime $k>0.15\hompc$,
where the Gaussian behaviour of the lognormal catalogues is expected
to break down. In fact we do find a turn-up of the data for
$k>0.15\hompc$, consistent with the non-linear behaviour of the
lognormal model. The top panel of Fig.~\ref{fig:Dsq_w1} shows the
average recovered power spectrum for $0.02<k<0.15\hompc$, calculated
from the mock catalogues weighted with weights $w_1$ (solid circles)
compared to the power expected from fitting to the radially averaged
Fourier window function for weighting scheme $w_1$ (dashed line); the
solid line is the input power spectrum (not convolved with the
window).

\begin{figure}
  \setlength{\epsfxsize}{\columnwidth} 
    \centerline{\epsfbox{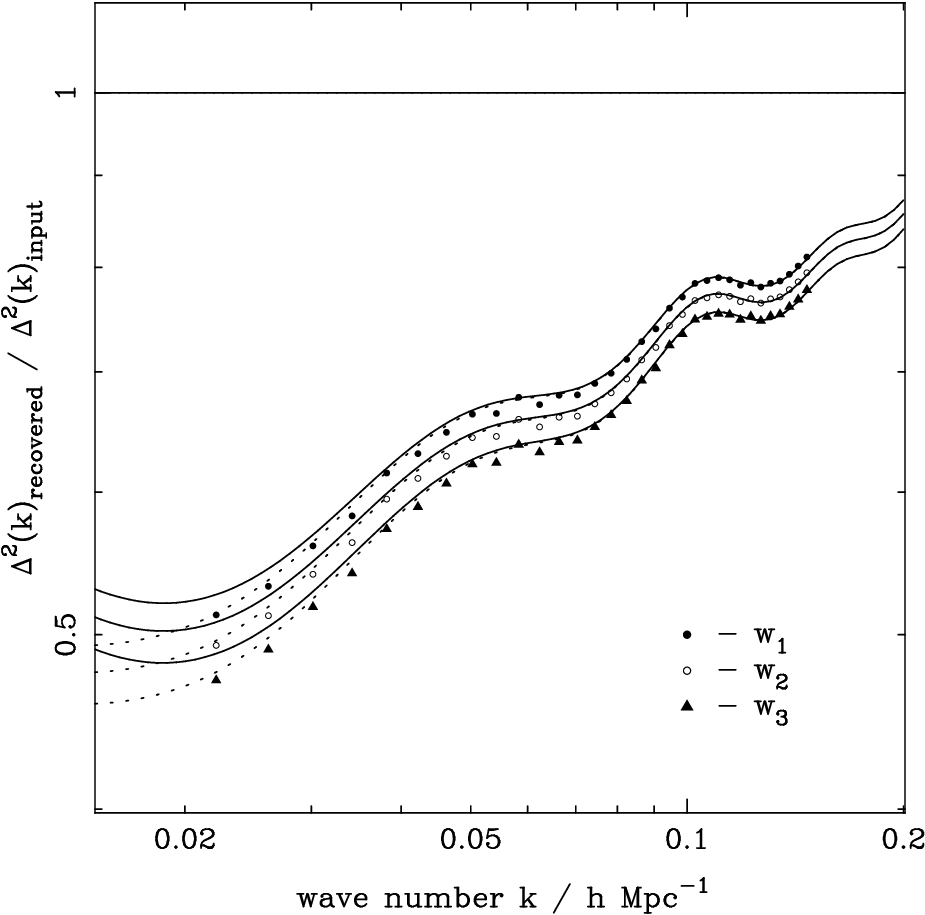}}

  \caption{The average recovered power spectrum from $1000$ mock
  catalogues, weighted as given by Eqns.~\ref{eq:w1}~to~\ref{eq:w3},
  ratioed to the input power (symbols). These data are compared with
  the expected power calculated by convolving a fit to the spherically
  averaged Fourier window function with the input power (solid
  lines). The dotted lines include the correction for the
  self-normalization induced by matching the average galaxy density
  (see Section~\ref{sec:alpha} for details).  \label{fig:pkconvw}}
\end{figure}

In order to examine the fit in detail and compare the average
recovered power from the three weighting schemes, we ratio this power
to the input power spectrum in Fig.~\ref{fig:pkconvw}. Here we see
that the increasing factor of bias in the weights $w_3\to w_1$
increases the effective size of the window in real space, and
therefore reduces the effect of the window in Fourier space: weighting
scheme $w_1$ produces the largest real-space volume as it increases
the importance of the rare high-luminosity galaxies at high redshift
the most.
 
The recovered diagonal elements of the covariance matrix calculated
using the three weighting schemes were very similar in amplitude, so
the number of realizations analysed was insufficient to distinguish
the relative errors in the recovered power produced by the different
the weighting schemes using this statistic. However, the smaller
$k$-space window function created using weights $w_1$ does mean that
this scheme is better at finding features in the power spectrum
compared with the other weighting schemes.

\begin{figure}
  \setlength{\epsfxsize}{\columnwidth} 
    \centerline{\epsfbox{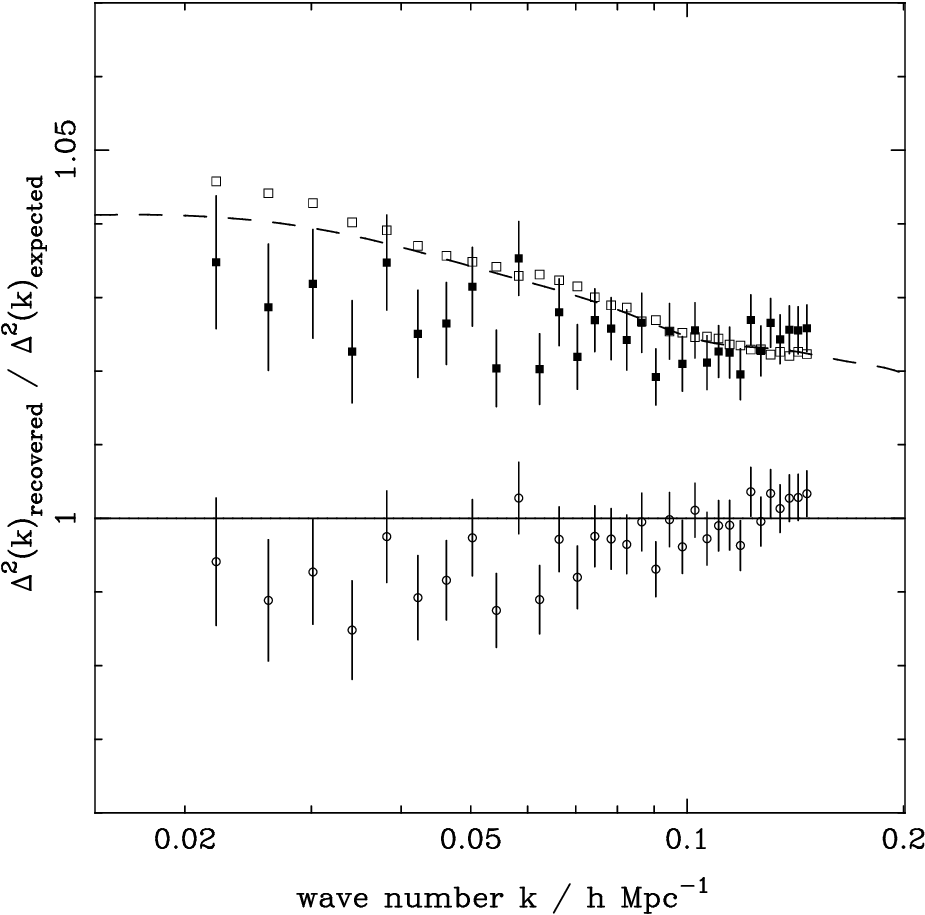}}

  \caption{The average recovered power from the mock catalogues,
  calculated using weight $w_2$ given by Eq.~\ref{eq:w2} ratioed to
  the expected power (open circles with 1$\sigma$ errors). The solid
  line shows the expected ratio of $1$. The solid squares with
  1$\sigma$ errors show the same average recovered data, compared to
  the model convolved with the window calculated from weights $w_3$,
  those assumed in the standard FKP analysis. The dashed line shows
  the expected ratio in this case. The errors on the data are
  dominated by cosmic variance: ratioing the two recovered average
  power spectra for the two weights gives the open squares. We see
  that not allowing for luminosity-dependent clustering, equivalent to
  assuming the wrong window, changes the power spectrum normalization
  and induces a slight change of slope.  \label{fig:2dFratio}}
\end{figure}

The difference between the convolved power for $w_2$ and $w_3$
corresponds to the small offset induced by not correcting for
luminosity-dependent clustering. This is illustrated in
Fig.~\ref{fig:2dFratio}, where the open circles show the ratio of the
average power spectrum recovered from the mock catalogues using the
original FKP method (calculated using weights $w_2$, Eq.~\ref{eq:w2})
with the expected power (the input power spectrum convolved with the
window function corresponding to weight $w_2$). The solid line show
the expected ratio of 1. The errors on the data are dominated by
cosmic variance and are given by the square root of the diagonal
elements of the covariance matrix. The solid squares with one sigma
errors show the ratio of the the same recovered power spectrum with
the matter power spectrum convolved with the window relative to
weights $w_3$, the window assumed without correcting for
luminosity-dependent clustering. The dashed line shows the expected
ratio. In the next section we consider how this offset affects the
2dFGRS analysis of P01.

\section{application to 2dFGRS analysis of P01}  \label{sec:2dFGRS}

\begin{figure}
  \setlength{\epsfxsize}{\columnwidth} 
    \centerline{\epsfbox{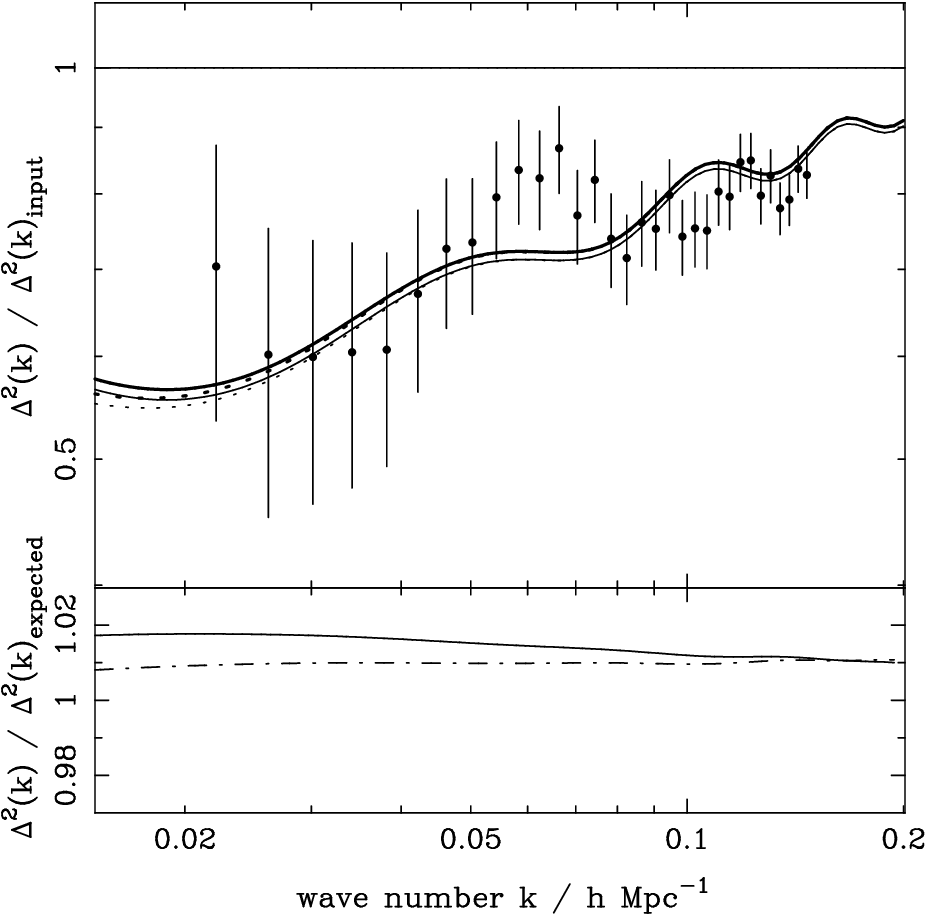}}

  \caption{Top panel: the 2dFGRS power spectrum data of P01 (solid
  points) with errors given by root of the diagonal elements of the
  covariance matrix. These data are compared with the concordance
  model ($\Omega_{\rm m}=0.3$, $\Omega_{\rm b}/\Omega_{\rm m}=0.15$,
  $\Omega_\Lambda=0.7$, $h=0.7$, $n_{\rm s}=1$) power spectrum
  convolved with the window function used in P01 (lower thin solid
  line), and with the corrected window function allowing for
  luminosity-dependent bias (upper thick solid line). In this plot,
  the 2dFGRS data have been fitted to the upper thick solid line. The
  dotted lines include the correction for the self-normalization of
  the galaxy number density. All of the data are ratioed to the
  unconvolved concordance model power spectrum. The difference between
  the two convolved models is negligible compared to the errors in the
  2dFGRS power. The ratio between the two convolved models gives the
  effect of not correcting for luminosity-dependent bias, and is
  plotted in the lower panel of this Figure (solid line). The change
  in shape induced is closely matched by a change in the spectral
  index of the fitted model -- the dot-dash line shows the ratio of a
  convolved power spectrum with $n=1$ and the window function allowing
  for luminosity dependent bias divided by a power spectrum with
  $n=0.995$ and a window function as assumed in a FKP analysis: the
  shape of the power spectrum recovered in a universe with $n=1$ would
  be fitted by a power spectrum with $n=0.995$ convolved with the
  original FKP window.  \label{fig:2dFGRS}}
\end{figure}

In the analysis of P01, no correction was made for the varying bias
caused by the range in galaxy luminosities within the sample. As
described in Section~\ref{sec:fkp_intro}, this causes the incorrect
window function to be assigned to the analysis. In the top panel of
Fig.~\ref{fig:2dFGRS} we compare the recovered 2dFGRS power spectrum
data with a model power spectrum convolved with both the window
function calculated using the FKP method, and that calculated after
correction for luminosity dependent bias (calculated with weights
given by $b(L)w(\r)$, see Section~\ref{sec:fkp_intro}). The corrected
window function has a smaller effect on the power compared with the
original because the extra bias weighting increases the effective size
of the survey. The models in this figure vary slightly from those
presented and used by P01, as we have fitted the spherically averaged
window functions using Spline3 fits as described in
Section~\ref{sec:mockmethod}. Changing how each spherically averaged
window was fitted has very little effect on the shape of the recovered
convolved power spectra, but it can affect the normalization ($\sim
10\%$), depending on the exact nature of the fits. This will be
discussed further in a subsequent paper.

The effect of not assuming the correct window when fitting the
recovered power spectrum slope and amplitude of the 2dFGRS analysis of
P01 is quantified in the lower panel of Fig.~\ref{fig:2dFGRS}.  The
solid line is the ratio of the expected power spectrum from a sample
of galaxies with luminosity-dependent bias within a concordance
universe ratioed to a theory power spectrum with the same cosmological
parameters but with $n=0.995$ instead of $n=1$, convolved with window
assumed in P01. This small shift in $n$ is sufficient to correct the
change in shape caused by assuming the wrong window, and is
approximately equivalent to a change $\Delta\Omega_mh\simeq0.0014$.
This change is obviously very small compared with the statistical
errors on the recovered parameters. There is also a small systematic
error on the power spectrum amplitude of $\simlt 2$\%, reflecting the
change in the normalization of the spherically averaged windows. The
solid line in the lower panel of Fig.~\ref{fig:2dFGRS} is comparable
with the dashed line in Fig.~\ref{fig:2dFratio}, which shows the same
effect for the mock catalogues. The difference between the two lines
is caused by the change in the denominator of Eqns.~\ref{eq:w2}
\&~\ref{eq:w3}, compared with the original FKP weightings. The extra
factor of $b^2(\r,L)$ in Eqns.~\ref{eq:w2} \&~\ref{eq:w3} decreases
the importance of the high redshift data, so the factor of $b(\r,L)$
on the numerator has more effect.

Using the revised window when fitting to the published 2dFGRS power
spectrum data of P01 changes the recovered $\Omega_mh$ and
$\Omega_b/\Omega_m$ values very slightly (to higher $\Omega_mh$ and
lower baryon fraction), but gives best-fit parameters
$\Omega_mh=0.20\pm0.03$ and $\Omega_b/\Omega_m=0.15\pm0.07$ (68\%
confidence interval, assuming $n_s=1$ \& $h=0.7\pm0.07$), that are the
same as those reported by P01 at this significance.

\section{discussion}  \label{sec:discussion}

We have presented a method for the Fourier analysis of galaxy redshift
surveys that allows for luminosity-dependent clustering.  Although
this generalization of the FKP analysis method is specifically
designed to allow for luminosity-dependent clustering, the derivation
is actually more general and can be applied to any mixed distribution
of galaxies with differing clustering strength. Consequently, it can
cope with type-dependent clustering, power spectrum evolution and bias
evolution, and should therefore be of use in the analysis of future
deeper redshift surveys. The method requires knowledge of the relative
change in the bias as a function of galaxy properties, although to
recover the matter power spectrum amplitude, the absolute bias is
required.

The optimal weights presented in this paper (Eq.~\ref{eq:w}) differ
from those of Yamamoto (2003), ignoring the correction for
redshift-space distortions (equation 29 of Yamamoto 2003). This
results from the fact that we included the effect of
luminosity-dependent clustering at the inception of our optimal weight
derivation and minimized the error in the underlying matter power
spectrum. Yamamoto (2003) instead minimized the error in the galaxy
power spectrum (biased power in the language of this
paper). Consequently in our weighting scheme, luminous galaxies are
given a higher weight than less luminous galaxies: they contribute
more signal-to-noise to the measure of the underlying fluctuations
than less luminous galaxies. However, this is not true if we wished to
optimize for the biased power as shown by Yamamoto (2003). The
weighting scheme that should be adopted depends on the statistic to be
measured: if we wish to measure the underlying matter power spectrum,
then the weights derived in Section~\ref{sec:optimise} are more
applicable.

The derivation of optimal weights was performed in the limit of a
negligible window function. On scales comparable to the size of the
window, the optimal weights will change, and will depend on the survey
geometry such that the weights no longer have radial symmetry
(e.g. Hamilton 1997). Although the three radially symmetric weighting
schemes considered in Section~\ref{sec:mockmethod} differed by a
factor of $\sim2$ over the redshift range of interest, they resulted
in remarkably similar diagonal errors in the power spectrum. If we
constrain the weights to be radially symmetric then small deviations
away from the derived optimal distribution do not appear to have a
significant effect on the recovered errors.

For P01, the error induced by not allowing for the effect of
luminosity-dependent clustering was shown to be negligible in
Section~\ref{sec:2dFGRS}. One of the factors that contributed to this
was that, in the weights applied, $P(k)$ was fixed at
$P(k)=5000\,h^3\,{\rm Mpc}^{-3}$. This was not the case in many other
Fourier analyses of galaxy redshift surveys (e.g. Tegmark, Hamilton \&
Xu 2002) which use a weight that varies as a function of $k$. Varying
the estimated power used in Eq.~\ref{eq:w} changes the normalization
$N$ (Eq.~\ref{eq:N}), which works in conjunction with the change in
shape of the spherically averaged window to alter the recovered
power. As we have shown in this paper, it is possible to correct these
effects by using the correct window function at each $k$-value.

Redshift space distortions caused by peculiar velocities will also
affect the shape of the recovered power spectrum. In the spirit of
generalizing FKP, such effects are not included here. However P01 used
mock catalogues drawn from the Hubble volume simulation to show that
these were not significant on the scales ($0.02<k<0.15\hompc$)
considered in this work. This issue will be explored further in a
later paper.

\section{Acknowledgments}
We thank Peder Norberg and Max Tegmark for stimulating
discussions. WJP is supported by PPARC through a Postdoctoral
Fellowship. LV is supported by NASA through Chandra Fellowship
PF2-30022 issued by the Chandra X-ray Observatory Center, which is
operated by the Smithsonian Astrophysical Observatory on behalf of
NASA under contract NAS8-39073. JAP is grateful for the support of a
PPARC Senior Research Fellowship. WJP and LV would like to thank the
Aspen Center for Physics where the initial stages of this work were
carried out during the ``Structure Formation in the Era of Large
Surveys'' workshop, June 2002.

\end{document}